# Molecular Dynamics Study of Droplets on Flat Crystalline Surface during Cooling and Ice formation

Yoshitaka Ueki[*,§], Yuta Tsutsumi[*] and Masahiko Shibahara[*]

[*]Department of Mechanical Engineering, Osaka University,
2-1 Yamadaoka, Suita, Osaka 565-0871, Japan
[§]Correspondence author.  Tel & Fax: +81-6-6879-4987
E-mail: ueki@mech.eng.osaka-u.ac.jp

**Abstract**

Condensation and frost formation degrade the heat transfer performance of air-conditioners and refrigerators. Yet, the frost formation mechanism has not been fully understood. In the present study, we numerically investigated $H_2O$ droplets during cooling and ice formation by means of classical molecular dynamics simulations. The mW potential was employed for the $H_2O$ molecules. A nanoscale $H_2O$ droplet was placed on a flat solid wall consisting of Pt atoms. The fcc (110) crystal plane faced the droplet. We examined where ice nucleation was formed and how the ice formation proceeded inside the droplet, and then evaluated temporal change in interfacial thermal resistance between the $H_2O$ molecules and the solid wall. We found that the cooling and ice formation changed the interfacial thermal resistance; however, its tendency differed depending on the solid-wall wettability. It was influenced by the nearest neighbor adsorption layer consisting of $H_2O$ molecules transformed during the cooling and ice formation.

## 1. Introduction

Degradation in the performances of air conditioners and refrigerators is caused by frost formation and adhesion on the heat transfer surface. The frost formation process is composed of a dropwise-condensation period, a solidified liquid tip-growth period, and a densification and bulk-growth period [1]. Through those periods, the frost branches form at the top of ice crystals, grow in multiple directions, and then get connected with neighboring frost branches, resulting in the formation of a flat frost layer. A previous experimental observation with X-ray μCT confirmed the above-mentioned frost formation process [2]. Another experimental observation showed that ice formation of a water droplet started from the cooling surface and then a pointy ice tip was formed on the top of the ice drop [3]. Some speculate that the frost branching growth on the top of the ice crystal was because initially volume expansion during water solidification forms the tip growth and then the ambient humid air induces condensation and solidification favorably on the ice tips. Conversely, some reported that the ice formation started from the gas-liquid interface of the water droplet, an ice shell was formed, and then the remaining liquid penetrated from the top of the ice shell as the inner liquid solidified. In spite of all the previous experimental studies, the frost formation mechanism has been not comprehensively understood yet.

Some molecular dynamics (MD) studies had been conducted to investigate ice nucleation and growth by means of classical MD simulations. One of the studies pinpointed that the crystal plane orientation of the ice that grew up from the heterogeneous ice nucleation depended on the lattice constant, the plane orientation of the crystalline surface, and the surface wettability [4]. In addition, the ice crystal can have multiple crystal plane orientations such as the case of ice Ih [5]. Since ice nucleation and growth are complicated phenomena as described above, the thermal energy transport in water droplets changes during the cooling and ice formation. On the molecular scale, interfacial thermal resistance (ITR) plays a significant role in thermal energy transport through solid-liquid and solid-solid interfaces. The ITR changes with its interface geometry [6,7], surface wettability [8], and other interface conditions. For example, previous MD studies showed that some surface ornamentations such as nanoparticle layers [9], graphene sheets [10], and self-assembled monolayers [11] on the solid-liquid interface influenced the ITR. Recently, some conducted a spectral analysis based on MD simulations and showed that phonon density of states and spectral heat flux could give further insight into the interfacial thermal transport [12-14].

Based on the background, in the present study by means of the classical MD method, we simulated how the water droplet cooled down and then the ice formed, and investigated how and how much the ITR between the water/ice droplet and the crystalline plane changed during both the cooling and ice formation period, with the different surface wettability.

## 2. Numerical Method

Figure 1 illustrates an example of the simulation system employed in the present study. The unit cell was 27.74 nm in the *x*-direction, 27.46 nm in the *y*-direction, and 16.96 nm in the *z*-direction. The periodic boundary condition was applied in the *x*- and *y*-directions. The mirror boundary condition was applied on the upper *z*-plane. The solid wall consisted of Pt atoms. The fluid molecules were $H_2O$. The ice nucleation rate depends on the plane orientations of the crystalline surface. Especially, the ice nucleations occur favorably on (110) and (111) surfaces of fcc structure [4]. Based on it, we employed the (110) surface of the fcc structure that faced the droplet. The total number of Pt atoms employed in the present simulation was 35000. That of the $H_2O$ molecules was 63000. The intermolecular potential for the $H_2O$ molecules was mW potential [15], which was expressed as follow:

$$\phi = \sum_i \sum_{j>i} \varphi_2(r_{ij}) + \sum_i \sum_{j \neq i} \sum_{k>j} \varphi_3(r_{ij}, r_{ik}, \theta_{ijk}) \quad (1)$$

$$\varphi_2(r_{ij}) = A\varepsilon \left[ B\left(\frac{\sigma}{r_{ij}}\right)^p - \left(\frac{\sigma}{r_{ij}}\right)^q \right] \exp\left(\frac{\sigma}{r_{ij} - a\sigma}\right) \quad (2)$$

$$\varphi_3(r_{ij}, r_{ik}, \theta_{ijk}) = \lambda\varepsilon \left[\cos\theta_{ijk} - \cos\theta_0\right]^2 \exp\left(\frac{\gamma\sigma}{r_{ij} - a\sigma}\right) \exp\left(\frac{\gamma\sigma}{r_{ik} - a\sigma}\right) \quad (3)$$

Table 1  mW potential parameters [15].

| $\varepsilon$  kcal/mol | $\sigma$  nm | $a$ | $\lambda$ | $\gamma$ |
|---|---|---|---|---|
| 6.189 | 0.23925 | 1.80 | 23.15 | 1.20 |
| $\cos\theta_0$ | $A$ | $B$ | $p$ | $q$ |
| −0.33333 | 7.04956 | 0.602225 | 4.0 | 0.0 |

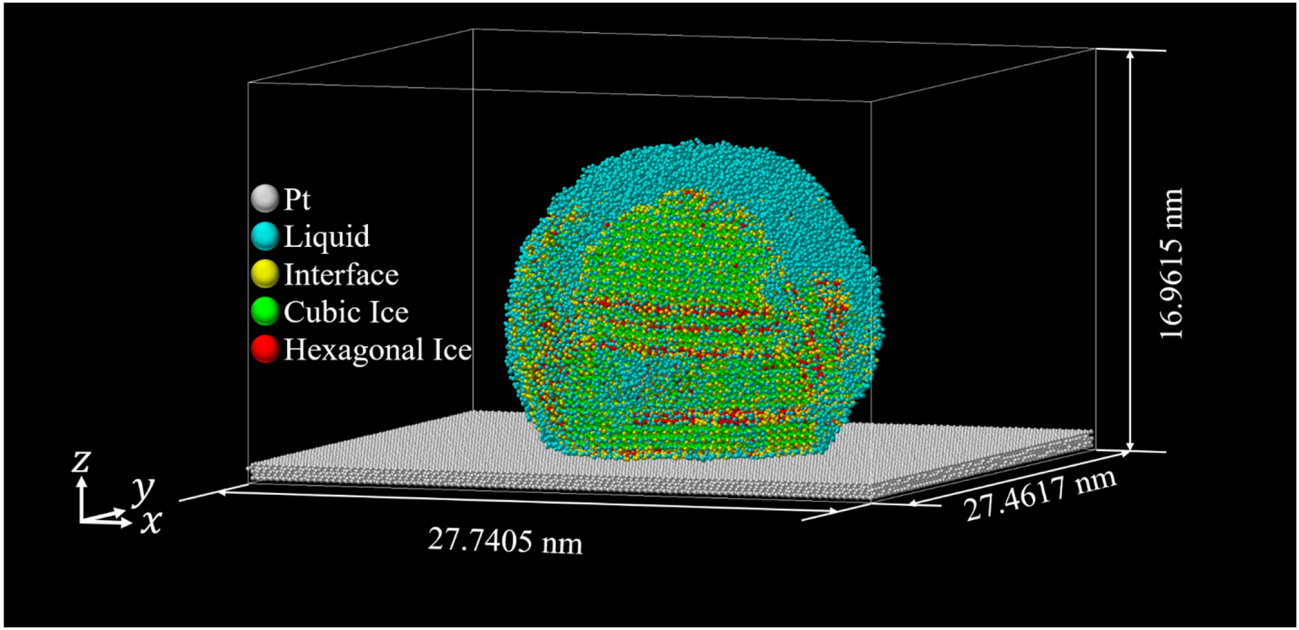

Figure 1. Simulation model.

The mW model is a coarse-grained $H_2O$ model and has been employed to investigate the $H_2O$ phase change phenomena and characteristics [16-18]. The mW potential parameters [15] were shown in Table 1. For the pairs of Pt atoms, we employed 12-6 Lennard-Jones intermolecular potential as expressed by Eq. (4). Moreover, for the pairs of the Pt atom and the $H_2O$ molecule, we employed the 12-6 Lennard-Jones potential multiplied with the interaction parameter $\alpha$, as expressed by Eq. (5).

$$\phi_{ij}(r_{ij}) = 4\varepsilon\left\{\left(\frac{\sigma}{r_{ij}}\right)^{12} - \left(\frac{\sigma}{r_{ij}}\right)^6\right\} \quad (4)$$

$$\phi_{ij}(r_{ij}) = 4\alpha\varepsilon\left\{\left(\frac{\sigma}{r_{ij}}\right)^{12} - \left(\frac{\sigma}{r_{ij}}\right)^6\right\} \quad (5)$$

Table 2 Lennard-Jones potential parameters [19].

|  | $\sigma$ nm | $\varepsilon$ J |
|---|---|---|
| Pt - Pt | 0.254 | $1.09 \times 10^{-19}$ |
| $H_2O$ - Pt | 0.2466 | $6.85 \times 10^{-20}$ |

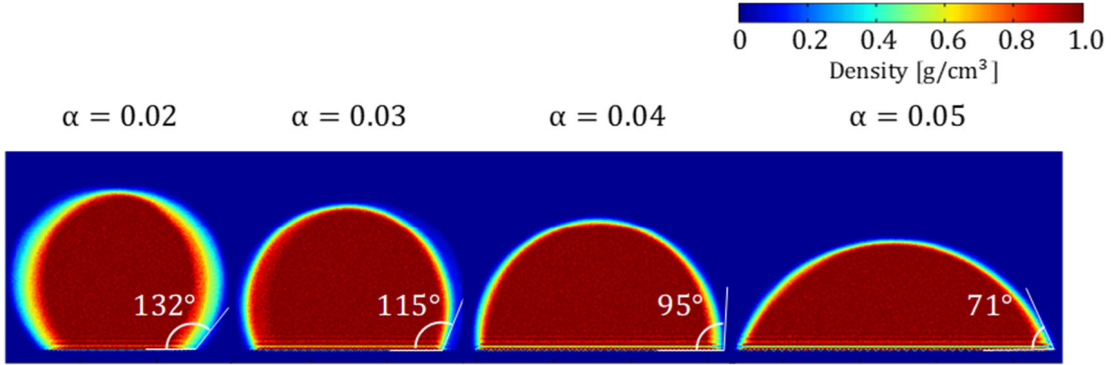

Figure 2. Contact angle at each surface wettability at 280 K.

The Lennard-Jones parameters employed in the present study [19] were shown in Table 2. The Lorentz-Berthelot combining rule was employed to determine the potential parameters of the pairs of Pt and $H_2O$. The numerical simulations were performed by LAMMPS [20]. Figure 2 illustrates contact angles of $H_2O$ on the Pt crystalline surface with different values of $\alpha$ at 280 K. The contact angle evaluation was based on the $\theta/2$ method.

The mass of the Pt atoms was 195.084 g/mol, and that of the $H_2O$ molecules was 18.01 g/mol. The time step of the numerical integration was 5.00 fs. The cut-off distance was 10 Å. The simulation procedures in the present study were as follows. We employed the Langevin method to control the temperature of the Pt walls, and the velocity scaling for the $H_2O$ molecules. In the first relaxation process of 2.5 ns, the $H_2O$ molecules and Pt atoms were maintained at 280 K by each of the thermosttaing methods. Then in the second relaxation process of 2.5 ns, the velocity scaling for the $H_2O$ molecules was removed. In the cooling and ice formation process, the Pt atoms were thermostatted at 205 K by the Langevin method.

We employed the CHILL algorithm to distinguish liquid from crystal [21]. The CHILL algorithm classifies the molecules based on the alignment of their orientation with respect to that of its four closest neighbors into the following group: cubic ice (Ic), hexagonal ice (Ih), interface, and liquid. Its classification coloring was shown in Figure 1. The ITR between the water/ice droplet and the Pt surface, $R$ was evaluated from the following equation:

$$R = \frac{\Delta T}{q} \qquad (6)$$

$\Delta T$ denotes the temperature difference between the closest neighbor adsorption layer of the water/ice droplet and the Pt surface. It was calculated from the time-averaged temperature of the water/ice droplet and the Pt surface. The time-averaging was 5 ps. The interrogation volume for the temperature calculation was the contact area of the water/ice droplet in the x- and y-directions, and the 5 Å from the contact plane in the z-direction. The interrogation area to determine if the water/ice droplet was in contact with the Pt surface was 5 Å x 5 Å in x- and y-directions. $q$ denotes the temporal heat flux of the thermal energy transferred by the Langevin method over the contact area of the water/ice droplet.

## 3. Results
### 3.1 Ice nucleation and growth

Figure 3 illustrates examples of the snapshots of the temporal cross-section of the water droplet during the ice formation in the case that $\alpha$ = 0.02 and 0.05. The white circles represent the ice nucleation points. In the present Pt-$H_2O$ system, the heterogeneous ice nucleation occurred on the Pt surface. In the case of a better wettability surface, more ice nucleation occurred at a faster rate. One of its reasons was that the droplet spread on the Pt surface and the contact area increased. In Figure 3, the white arrows inside the droplets represent the ice formation direction. The ice formation proceeded from the Pt surface toward the droplet tip. The ice structure formed in the present study was a laminate structure of the hexagonal and cubic ices. In the case of the relatively poor wettability of $\alpha$ = 0.02, the laminate structure grew up in the off-vertical direction with respect to the Pt surface. It was because not much ice nucleation happened on the poor wettability surface. In the case of the relatively better wettability surface, simultaneously multiple ice nucleation happened. In the ice growing period, they merged and then grew up in the vertical direction.

Figure 4 illustrates the time change in the fraction of the ice, which was the cubic, hexagonal, and interface, at different values of $\alpha$. It shows that, as the Pt surface had better wettability, the ice nucleation occurred earlier and the ice formation faster. In addition, note that the saturated ice fraction was approximately 0.9 and the remaining part was in a form of a liquid or amorphous state. Figure 5 shows the snapshots of the cross-section of the ice droplets in the steady-state in the case that $\alpha$ = 0.02 and 0.05. From Figure 5, there were some liquid/amorphous molecules inside the ice droplets. The quasi-liquid state, where the water molecules stayed melted even under the solidification temperature, could be seen on the ice interface [22]. Furthermore, the amorphous ice could form from the supercooled water [23]. The quasi-liquid layer or/and the amorphous ice formed inside the ice droplet, and so the ice droplet were not completely in the crystalline forms.

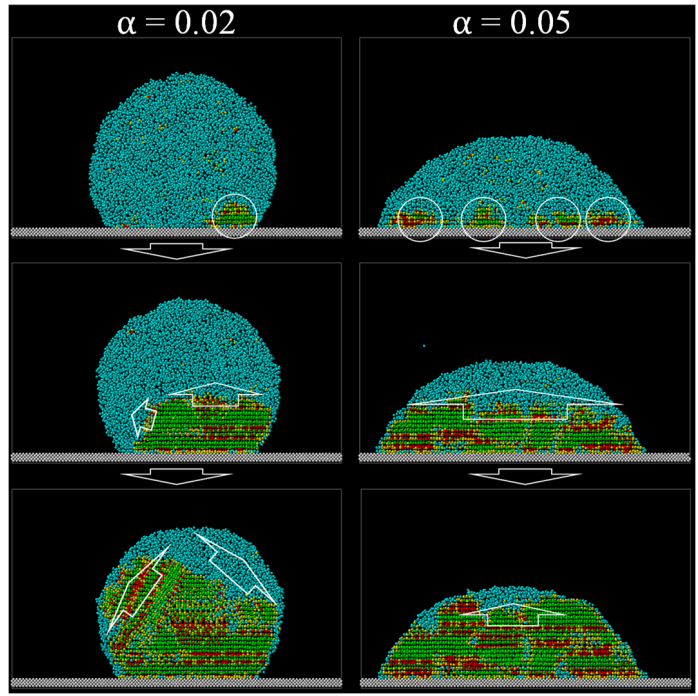

Figure 3. Snapshots of temporal cross-section of water droplet during ice formation; molecules coloring was based on distinction of CHILL algorithm.

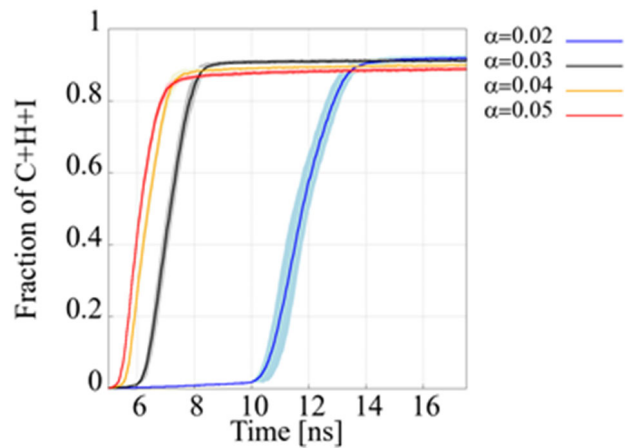

Figure 4. Time change in the fraction of ice, which was cubic, hexagonal, and interface, at different values of $\alpha$; higher values of $\alpha$ corresponds to better surface wettability.

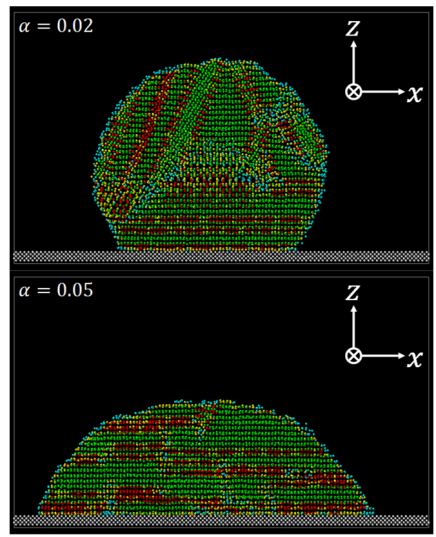

Figure 5. Snapshots of cross-section of ice droplets in steady-state in the case of $\alpha = 0.02$ and $0.05$; molecules coloring was based on distinction of CHILL algorithm.

## 3.2 Interfacial Thermal Resistance

Figure 6 illustrates the time change in the ITR and the ice fraction at different values of $\alpha$. For each value of $\alpha$, we performed the above-mentioned simulations five times. Those averaged values and error bars were shown in Figure 6. The standard variance was shown as the error bars. In the present study, the order of magnitude of the ITR was $10^{-8} – 10^{-9}$ m²K/W. It was consistent with a previous study of the ITR between the mW $H_2O$ molecules and the Lennard-Jones Pt surface [24-25]. In the study, the order of magnitude of the ITR was $10^{-8}$ m²K/W. Note that the plane orientation of the Pt crystalline surface and some of the other conditions were not identical. According to the other previous studies, the order of magnitude of the solid-liquid ITR was $10^{-9}$ - $10^{-7}$ m²K/W, depending on the definitions of the thermal resistance, the surface wettability, and the fluid molecular models [e.g., 6-14, 26-28]. The order of magnitude of the ITR in the present study was consistent with the above-mentioned previous studies.

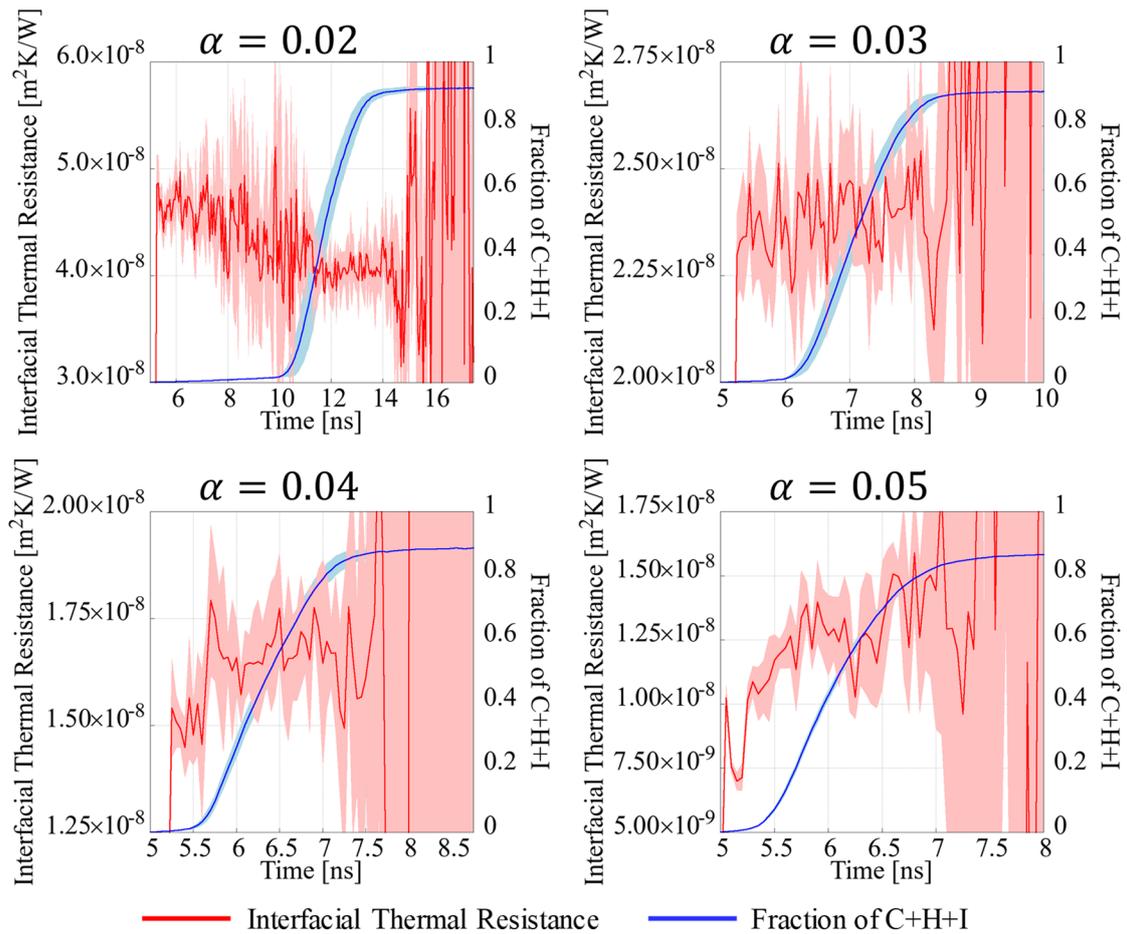

Figure 6. Time change in ITR and ice fraction at different value of $\alpha$.

In the case of $\alpha$ = 0.02, the ITR decreased during the cooling and ice formation. And, in the case of $\alpha$ = 0.03, the ITR did not significantly change during the relevant processes. On the other hand, in the cases of $\alpha$ = 0.04 and 0.05, the ITR increased during them. It showed that the ITR during the cooling and ice formation significantly depended on the wettability of the (110) crystalline plane surface.

Figure 7 shows the time change in the $H_2O$ molecules density distribution of the closest neighbor adsorption layer in the *z*-direction at the different values of $\alpha$. The density distributions were calculated by averaging five simulation runs under identical conditions. It showed that, in the case of $\alpha$ = 0.02, the density of the $H_2O$ molecules was increasing in the vicinity of the Pt surface during the cooling and ice formation, and formed the twin peak of the density distribution due to the ice crystalization. It meant that, as an overall trend, the $H_2O$ molecules approached the Pt surface, and as a result, the thermal energy transport augmented due to the more active intermolecular interaction across the Pt-$H_2O$ interface. It caused that the ITR was decreasing during the relevant processes.

Also in the case of $\alpha$ = 0.03, Figure 7 showed that, as the overall trend, the $H_2O$ molecules approached the Pt surface. However, Figure 7 (b) showed that the density distribution, where $z$ = 0.85 – 0.90 nm, decreased after the time of 5.75 ns. It meant that the closest neighbor peak went far off from the Pt surface. The previous MD study [24-25] showed that the proximity $H_2O$ molecules, which were located between the Pt surface and the closest density peak, transported a major quantity of the thermal energy across the Pt-$H_2O$ interface. Based on them, the positive effect of the change in the density distribution and the negative effect of the fewer proximity molecules balanced out, resulting in the approximately steady ITR during the cooling and ice formation.

In the cases of $\alpha$ = 0.04 and 0.05, similarly, the density distribution transformed due to the ice formation, and the proximity molecules became less. From Figure 7 (b), it is notable that, at $\alpha$ = 0.04, the proximity molecules became less after the time of 5.45 ns and that, at $\alpha$ = 0.05, they did after the time of 5.30 ns. In addition, Figure 7 showed that the closest neighbor peak became higher, as the surface wettability improved. In the cases, the negative effect of the fewer proximity molecules outweighed the above-mentioned positive effect. As a result, the ITR increased during the relevant processes.

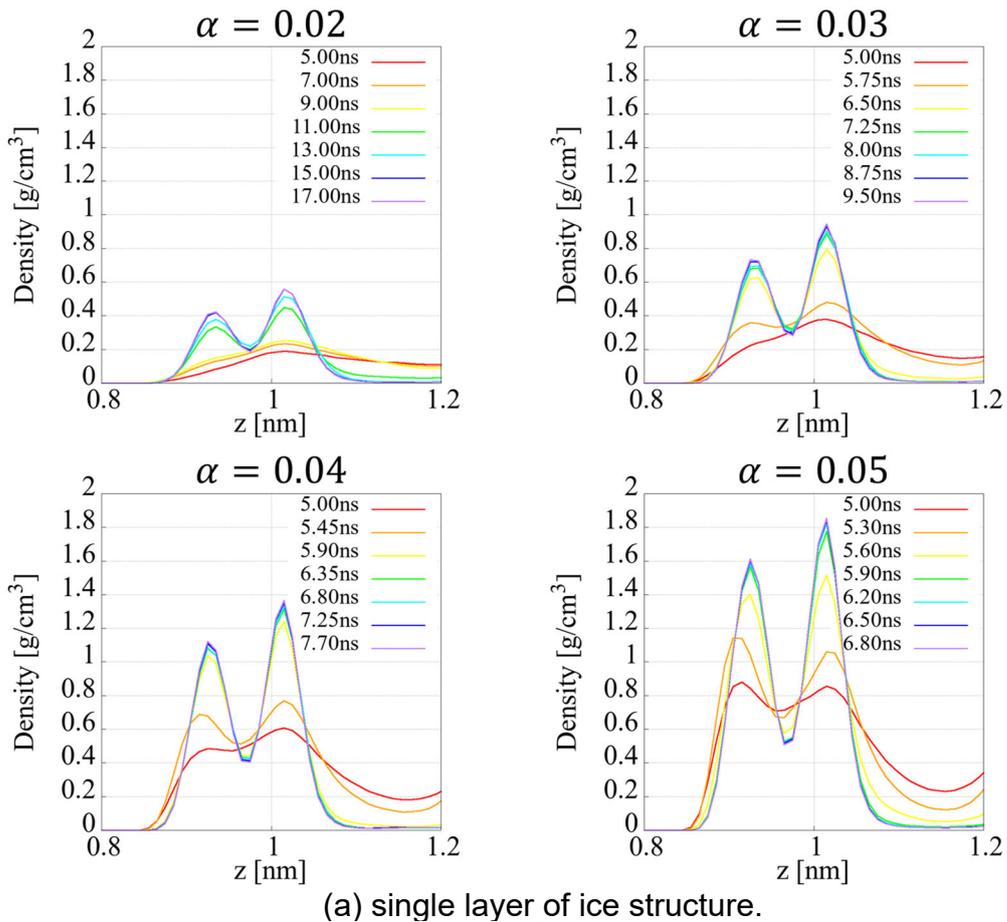

(a) single layer of ice structure.

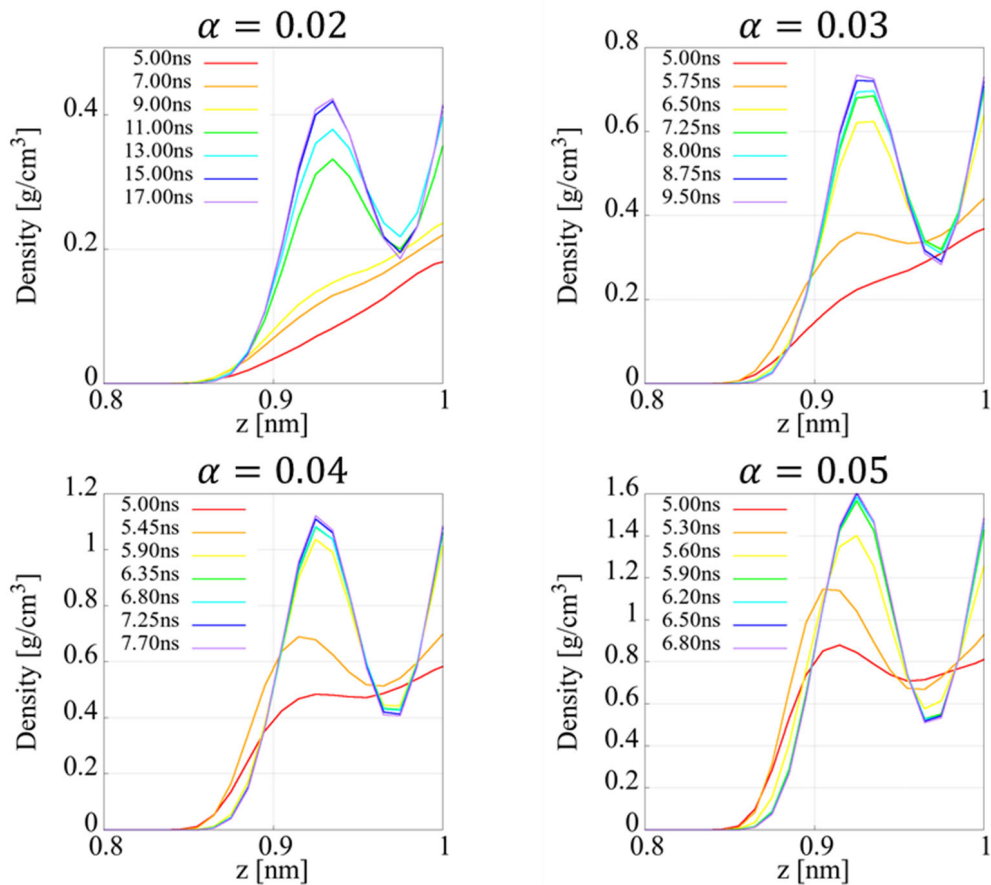

(b) enlarged illustration.

Figure 7. Time change in $H_2O$ molecules density distribution of closest neighbor adsorption layer in z-direction at different value of $\alpha$.

## Conclusions

In the present study, by means of the classical non-equilibrium MD simulation, we investigate how and how much the ITR between the water/ice droplet and the Pt crystalline plane changed during both the cooling and ice formation period, with the different surface wettability. Our findings are summarized as follows:

- In the present Pt-$H_2O$ system, the homogeneous ice nucleation occurred favorably on the Pt surface, and the ice formation proceeded toward the droplet tip. On the relatively better wettability surface, the multiple ice nucleation occurred early and simultaneously. On the relatively poor wettability surface, biased ice formation happened.
- The change in the ITR between the water/ice droplet and the crystalline Pt surface depended on the surface wettability.
- The main factor of the change in the ITR during the cooling and ice formation was the transformation of the closest neighbor structure of the $H_2O$ molecules. In addition to it, we speculated that the proximity molecules, which were located between the Pt surface and the closest density peak, played an important role to transport the thermal energy across the Pt-$H_2O$ interface, and that these factors determined the ITR.

## Nomenclature

| | |
|---|---|
| $i, j, k$ | molecular index |
| $q$ | heat flux |
| $R$ | interfacial thermal resistance |
| $r$ | intermolecular distance |
| $\alpha$ | interaction parameter |
| $\Delta T$ | temperature difference |
| $\varepsilon$ | depth of intermolecuar potential well |
| $\theta_{ijk}$ | bonding angle of molecular $i, j, k$ |
| $\sigma$ | apparent molecular radius |
| $\phi$ | intermolecular potential |

## Acknowledgment

This work was supported by JSPS KAKENHI Grant Number 18H01382.

# References

[1] M. Song, C. Dang, "Review on the measurement and calculation of frost characteristics", International Journal of Heat and Mass Transfer, **124**, pp. 586-614 (2018).

[2] R. Matsumoto, T. Uechi, Y. Nagasawa, "Three-dimensional microstructure of frost layer measured by using X-ray μCT", Journal of Thermal Science and Technology, **13(1)**, (2018).

[3] A. G. Marín, O. R. Enríquez, P. Brunet, P. Colinet, J. H. Snoeijer, "Universality of tip singularity formation in freezing water drops", Physical Review Letters, **113,** 054301 (2014).

[4] M. Fitzner, G. C. Sosso, S. J. Cox, A. Michaelides, "The many faces of heterogeneous ice nucleation: interplay between surface morphology and hydrophobicity", Journal of American Chemical Society, **137**, 42, 13658–13669 (2015).

[5] J. A. Hayward, A. D. J. Haymet, "The ice/water interface: Molecular dynamics simulations of the basal, prism, $\{20\bar{2}1\}$, and $\{2\bar{1}\bar{1}0\}$ interfaces of ice Ih", Journal of Chemical Physics, **114**(8), 3717-3726 (2001).

[6] Y. Wang, P. Kebliski, "Role of wetting and nanoscale roughness on thermal conductance at liquid-solid interface", Applied Physics Letters, **99** 073112 (2011).

[7] M. Shibahara et al., "Molecular Dynamics Study on Influences of Surface Structural Characteristics on Thermal Energy Transport over Liquid-Solid Interfaces", International Heat Transfer Conference Digital Library (2014). DOI: 10.1615/IHTC15.mlt.008513

[8] Z. Ge et al., "Thermal Conductance of Hydrophilic and Hydrophobic Interfaces", Physical Review Letters, 96, 186101 (2006).

[9] Y. Ueki et al., "Molecular dynamics study of thermal resistance of solid-liquid interface in contact with single layer of nanoparticles", International Journal of Heat and Mass Transfer, **120,** 608-623 (2018).

[10] A. T. Pham et al., "Interfacial thermal resistance between the graphene-coated copper and liquid water", International Journal of Heat and Mass Transfer, **97**, 422-431 (2016).

[11] G. Kikugawa, "A molecular dynamics study on heat transfer characteristics at the interfaces of alkanethiolate self-assembled monolayer and organic solvent", Journal of Chemical Physics, **130**, 074706 (2009).

[12] A. Giri, P. E. Hopkins, "Spectral analysis of thermal boundary conductance across solid/classical liquid interfaces: A molecular dynamics study", Applied Physics Letters, **105,** 033106 (2014).

[13] K. Sääskilahti et al., "Spectral mapping of heat transfer mechanisms at liquid-solid interfaces", Physical Review E, **93**, 052141 (2016).

[14] B. Ramos-Alvarado, S. Kumar, "Spectral Analysis of the Heat Flow Across Crystalline and Amorphous Si–Water Interfaces", The Journal of Physical Chemistry C, **121**(21) 11380-11389 (2017).

[15] V. Molinero et al., "Water Modeled As an Intermediate Element between Carbon and Silicon", Journal of Physical Chemistry B, **113**(13) (2009) 4008-4016.

[16] M. Fitzner et al., "Predicting heterogeneous ice nucleation with a data-driven approach", Nature Communications, 11:4777 (2020).

[17] R. Shi and H. Tanaka, "The anomalies and criticality of liquid water", PNAS, **117** (43) 26591-26599 (2020).

[18] L. Lupi et al., "Role of stacking disorder in ice nucleation", Nature, **551**, 218–222 (2017).

[19] S. B. Zhu et al., "Interaction of water with metal surfaces", The Journal of Chemical Physics, **100**(9), 6961-6968 (1994).

[20] S. Plimpton, "Fast Parallel Algorithms for Short-Range Molecular Dynamics", Journal of Computational Physics, **117**, 1-19 (1995).

[21] E. B. Moore et al., "Freezing, melting and structure of ice in a hydrophilic nanopore", Physical Chemistry Chemical Physics, **12**, 4124-4134 (2010).

[22] H. Asakawa et al., "Two types of quasi-liquid layers on ice crystals are formed kinetically", PNAS, **113** (7), (2016), pp. 1749-1753.

[23] E. B. Moore and V. Molinero, "Is it cubic? Ice crystallization from deeply supercooled water", Physical Chemistry Chemical Physics, **13**, (2011), pp. 20008-20016.

[24] Y. Ueki, S. Matsuo, M. Shibahara, "Molecular Dynamic Study of Local Interfacial Thermal Resistance of Solid-Liquid and Solid-Solid Interfaces: Water and Nanotextured Surface", arXiv preprint, arXiv:2011.03184 (2020).

[25] Y. Ueki, S. Matsuo, M. Shibahara, "Molecular Dynamic Study of Local Interfacial Thermal Resistance of Solid-Liquid and Solid-Solid Interfaces: Water and Nanotextured Surface", Proceedings of 8th International Symposium on Advances in Computational Heat Transfer, CHT-21-135, (2021).

[26] L. Hu et al., "Determination of interfacial thermal resistance at the nanoscale", Physical Review B, **83**, 195423 (2011).

[27] Z. Liang et al., "Thermal resistance at an interface between a crystal and its melt", The Journal of Chemical Physics, **141**, 014706 (2014).

[28] D. Surblys et al., "Molecular dynamics investigation of surface roughness scale effect on interfacial thermal conductance at solid-liquid interfaces", The Journal of Chemical Physics, **150**, 114705 (2019).